\documentclass[11pt]{amsart}

\usepackage{amsmath, amscd,hyperref,amssymb}
\usepackage{amsfonts}
\usepackage{amsthm, upref}
\usepackage{graphicx}
\usepackage{enumerate}
\usepackage[dvipsnames]{xcolor}
\usepackage{tikz}
\usepackage{pgf}
\usetikzlibrary{matrix,arrows, decorations.markings}
\usepackage{accents}

\input xy
\xyoption{all}

    \numberwithin{equation}{section}

\usepackage{bm}
        
\bmdefine\alphab{\mathbf{\alpha}}
\bmdefine\betab{\mathbf{\beta}}
\bmdefine\pib{\mathbf{\pi}}
\bmdefine\xib{\mathbf{\xi}}
\bmdefine\sigmab{\mathbf{\sigma}}

\newcommand{\comment}[1]{}
\newcommand{\eq}{\begin{equation}}
\newcommand{\en}{\end{equation}}

\newcommand{\rr}{\mathbb{R}}

\newcommand{\ep}{\hfill $\Box$}

\begin{document}

\theoremstyle{plain}
\newtheorem{thm}{Theorem}[section]
\newtheorem{rem}{Remark}[section]
\newtheorem{lemma}[thm]{Lemma}
\newtheorem{prop}[thm]{Proposition}
\newtheorem{cor}[thm]{Corollary}

\theoremstyle{definition}
\newtheorem{defn}{Definition}[section]
\newtheorem{cond}{Condition}
\newtheorem{asmp}{Assumption}
\newtheorem{notn}{Notation}
\newtheorem{prb}{Problem}

\theoremstyle{remark}
\newtheorem{rmk}{Remark}[section]
\newtheorem{exm}{Example}[section]
\newtheorem{clm}{Claim}

\numberwithin{equation}{section}

\newcommand{\ms}[1]{\color{RedOrange} {#1}}

\title[Relative arbitrage problem]{Relative arbitrage problem under eigenvalue lower bounds}

\author{Jou-Hua Lai, Mykhaylo Shkolnikov and H.~Mete Soner}
\address{ORFE Department \\ Princeton University \\ Princeton, NJ 08544}
\email{jhlai@princeton.edu }
\address{Department of Mathematical Sciences and Center for Nonlinear Analysis\\Carnegie Mellon University \\  Pittsburgh, PA 15213}
\email{mshkolni@gmail.com}
\address{ORFE Department \\ Princeton University \\ Princeton, NJ 08544}
\email{soner@princeton.edu}

\keywords{Mean curvature flow, nonlinear elliptic PDE, portfolio domination, relative arbitrage, stochastic optimal control, stochastic portfolio theory, sufficient volatility, viscosity solutions}  

\subjclass[2020]{91G10, 93E20, 49L25, 53E10}

\thanks{M.~Shkolnikov is partially supported by the National Science Foundation grant DMS-2342349. H.~M.~Soner is partially supported by the National Science Foundation grant DMS-2406762.}

\date{\today}

\begin{abstract}
We give a new formulation of the relative arbitrage problem from stochastic portfolio theory that asks for a time horizon beyond which arbitrage relative to the market exists in all ``sufficiently volatile'' markets. In our formulation, ``sufficiently volatile'' is interpreted as a lower bound on an ordered eigenvalue of the instantaneous covariation matrix, a quantity that has been studied extensively in the empirical finance literature. Upon framing the problem in the language of stochastic optimal control, we characterize the time horizon in question through the unique upper semicontinuous viscosity solution of a fully nonlinear elliptic partial differential equation (PDE). In a special case, this PDE amounts to the arrival time formulation of the Ambrosio-Soner co-dimension mean curvature flow. Beyond the setting of stochastic portfolio theory, the stochastic optimal control problem is analyzed for arbitrary compact, possibly non-convex, domains, thanks to a boundedness assumption on the instantaneous covariation matrix. 
\end{abstract}

\maketitle

\section{Introduction} \label{sec1}

In \cite[Section 3.3]{fernholz_stochastic_2002}, Fernholz has introduced the concept of portfolio domination, now more commonly referred to as \textit{relative arbitrage}, to rigorously capture the notion of ``beating the market''. More specifically, consider a market with $d\ge2$ assets whose vector $\mu:=(\mu_1,\mu_2,\ldots,\mu_d)$ of market weights (i.e., market capitalizations as fractions of the total market capitalization) constitutes a continuous semimartingale with respect to some stochastic basis. A predictable $\mu$-integrable process $\theta$ is called a trading strategy. We refer to $V^{\theta}:=\theta^{\top} \mu := \sum_{i=1}^d \theta_i \,\mu_i$ as the value process of $\theta$ relative to the market and only consider self-financing trading strategies, meaning:  
\begin{equation}
V^{\theta}(t) = V^\theta(0)+\int_0^t \theta(s)^{\top}\,\mathrm{d}\mu(s), \quad t\geq 0.
\end{equation}
Taking the market (i.e., the trading strategy $(1,1,\ldots,1)$) as the benchmark, we can state the definition of relative arbitrage from the seminal monograph by Fernholz and Karatzas (\cite[Definition 6.1]{karatzas_stochastic_2009}) as follows.

\begin{defn}
A trading strategy $\theta$ is called a \textit{relative arbitrage} over a time horizon $[0, T]$ if its value process $V^{\theta}$ relative to the market satisfies:
\begin{enumerate}[(a)]
    \item $V^\theta\ge0$ almost surely, 
 	\item $V^{\theta}(T) \ge V^{\theta}(0)$ almost surely,
    \item $V^{\theta}(T) > V^{\theta}(0) $ with positive probability.
\end{enumerate}
\end{defn}

\smallskip

The terms `trading strategy', `value process', `self-financing', and `arbitrage' used above
take on their customary meaning when one starts from a market without a bank account and employs the value of the portfolio continuously investing according to the market weights (`market portfolio') as the num\'{e}raire. The latter is emphasized by the term `relative (to the market)'. 

\medskip

A central problem of stochastic portfolio theory, going back to the foundational work \cite{fernholz_stochastic_2002} by Fernholz, is to identify classes of markets that admit relative arbitrages over suitable time horizons $[0,T]$. In particular, it is proven in \cite[Example 3.3.3]{fernholz_stochastic_2002} that every market in which the smallest eigenvalue of the underlying assets' instantaneous covariation matrix is bounded away from $0$ and the largest market weight is bounded away from $1$ admits a relative arbitrage over all long enough time horizons $[0,T]$. Surprisingly, the existence of relative arbitrages for the same kind of markets has been shown to hold over \textit{any} non-trivial time horizon in \cite{fernholz_diversity_2005}. This phenomenon, now known as \textit{short-term relative arbitrage}, has been demonstrated in \cite{fernholz_relative_2005} also for the so-called volatility-stabilized markets. We refer further to \cite{KR17}, \cite{vovk18}, \cite{Cu19}, \cite{RX19}, \cite{KK20}, \cite{itkin25} for various extensions of these findings. 

\medskip

The practically important distinction between relative arbitrage over long enough time horizons and short-term relative arbitrage naturally prompted the following question (see \cite[Section 4]{BF08}). Does every sufficiently volatile market, in the sense of 
\begin{equation}\label{FK cond}
\sum_{i=1}^d  \frac{1}{\mu_i(t)}\,\frac{\mathrm{d}{\langle\mu_i\rangle}(t)}{\mathrm{d}t}  \ge C>0,\quad t\geq 0
\end{equation}
-- a variant of the key assumption behind the construction in \cite[Example 3.3.3]{fernholz_stochastic_2002} highlighted in \cite[Section 3]{fernholz_relative_2005}\footnote{Note that the left-hand side in \eqref{FK cond} is precisely the `excess growth rate' in \cite[display (3.1)]{fernholz_relative_2005}.}, admit a short-term relative arbitrage, and not only relative arbitrages over long enough time horizons? In another surprising twist, this question has been answered negatively in  \cite[Section 6]{fernholz_volatility_2018}. Hence, the focus has shifted to finding the \textit{smallest} $T^*>0$ such that a relative arbitrage over the time horizon $[0,T]$ is possible for any $T>T^*$ in every sufficiently volatile market, usually referred to as the \textit{relative arbitrage problem}.      

\medskip

The relative arbitrage problem appears intractable at first glance, but two remarkable insights by Larsson and Ruf (see \cite{larsson_relative_2021}) have allowed them to characterize $T^*$ when the sufficient volatility condition \eqref{FK cond} is replaced by 
\begin{equation}\label{LR cond}
\sum_{i=1}^d \langle\mu_i\rangle(t)\ge t,\quad t\geq 0
\end{equation}
(a prominent variant of \eqref{FK cond}, see, e.g., \cite[Example 5.5]{KR17} and apply a time change). Firstly, \cite{larsson_relative_2021} uses the Fundamental Theorem of Asset Pricing to express $T^*$ through the value function of a stochastic optimal control problem. Secondly, \cite{larsson_relative_2021} identifies the Hamilton-Jacobi-Bellman equation associated with the latter as the arrival time formulation of the minimum curvature flow, a geometric flow akin to the celebrated mean curvature flow (see \cite{Mullins56}, \cite{Brakke78}, \cite{Huisken84}), on the probability simplex. The arrival time formulation of the minimum curvature flow on the probability simplex turns out to have a unique viscosity solution that eventually characterizes $T^*$ (see \cite[Theorem~5.1]{larsson_relative_2021}). This resolves the relative arbitrage problem under the sufficient volatility condition \eqref{LR cond}.

\medskip

In the setting of \cite{larsson_relative_2021}, the process of market weights $\mu$ 
free of relative arbitrage over the time horizon $[0,T^*]$ is expected to have an instantaneous covariation matrix $\big(\frac{\mathrm{d}\langle \mu_i,\mu_j\rangle(t)}{\mathrm{d}t}\big)_{1\le i,j\le d}$ generically of rank $1$  (cf.~\cite[discussion following Remark 1.4]{larsson_minimum_2022}). At the same time, while instantaneous covariation matrices of large asset universes are commonly estimated with factor models in the empirical finance literature (see \cite[Section 8]{johansson_simple_2023} for a concise overview) and the number of factors used -- equal to the number of dominant eigenvalues reliably estimated -- is indeed much smaller than $d$, this number tends to be much larger than $1$ (e.g., $20$ for $d=238$ in \cite[Figure 8.1]{johansson_simple_2023}). For this reason, we amend the sufficient volatility condition \eqref{LR cond} to 
\begin{equation}\label{LSS cond}
\lambda_{(n-k)}\bigg(\frac{\mathrm{d}\langle \mu_i,\mu_j\rangle(t)}{\mathrm{d}t}\bigg)_{1\le i,j\le d}\ge 1,\quad \text{a.e. }t\ge0
\end{equation}
where $n:=d-1$, $k$ is any fixed element of $\{1,2,\ldots,n-1\}$, and $\lambda_{(n-k)}$ refers to the $(n-k)$-largest eigenvalue. A situation such as in \cite[Figure 8.1]{johansson_simple_2023} then has $n=237$ and $k=217$ for example. To obtain a positive constant other than $1$ on the right-hand side of~\eqref{LSS cond}, it suffices to apply a simple time change throughout. 

\medskip

The ``null hypothesis'' \eqref{LSS cond} is not closed with respect to the convergence in distribution of continuous processes because the lower bound of \eqref{LSS cond} is not preserved under convex combinations of instantaneous covariation matrices. Therefore, we convexify \eqref{LSS cond} (cf. Lemma \ref{lem: convex hull} below): 
\begin{equation}\label{LSS cond'}
\Pi_m\bigg(\frac{\mathrm{d}\langle \mu_i,\mu_j\rangle(t)}{\mathrm{d}t}\bigg)_{1\le i,j\le d}\geq m-k \quad\text{for}\quad m = k+1,\, k+2,\,\ldots,\, n,\quad \text{a.e. } t\ge0,
\end{equation}
where $\Pi_m(a):= \inf\{\text{tr}(aP)\!: P^2 = P,\,\text{tr}(P)=m\}$ and $\text{tr}$ stands for the trace of a square matrix. Hence, we study the relative arbitrage problem under the sufficient volatility condition \eqref{LSS cond'} together with the technical condition 
$\lambda_{(1)}\big(\frac{\mathrm{d}\langle \mu_i,\mu_j\rangle(t)}{\mathrm{d}t}\big)_{1\le i,j\le d}\le L$, $\text{a.e. }t\ge0$ for some fixed $L\ge1$. The latter may be chosen large enough to accommodate a confidence interval around an empirical estimate of $\lambda_{(1)}\big(\frac{\mathrm{d}\langle \mu_i,\mu_j\rangle}{\mathrm{d}t}\big)_{1\le i,j\le d}$. The compactness of the resulting set of instantaneous covariation matrices allows us to establish the semicontinuity of the value function and the dynamic programming principle in the ensuing stochastic optimal control problem. 

\medskip

To characterize $T^*$, we begin as in \cite[Section 5]{larsson_relative_2021}. More specifically, we apply a linear transformation $U$ mapping the probability simplex isometrically onto a polytope $K\subset\rr^n$. Then, an application of the Fundamental Theorem of Asset Pricing as in \cite[proof of Theorem 3.1]{larsson_relative_2021} yields the representation $T^*=v(U\mu(0))$, where  
\begin{equation}\label{soc problem}
v(x) := \sup_{\mathrm{P} \in \mathcal{P}_x} \mathrm{P}\text{-ess} \inf \tau_K\,;
\end{equation}
$\mathcal{P}_x$ is the set of probability measures on $\Omega:=C([0,\infty),\mathbb{R}^n)$, equipped with the Borel $\sigma$-algebra for the topology of locally uniform convergence, under which the coordinate process $X$ is a martingale starting from $x$ and
\begin{equation}\label{eq: constraint}
\begin{split}
& \Pi_m\bigg(\frac{\mathrm{d}\langle X_i,X_j\rangle(t)}{\mathrm{d}t}\bigg)_{1\le i,j\le n}\geq m-k \quad\text{for}\quad m = k+1,\, k+2,\,\ldots,\, n,\quad \text{a.e. }t\ge0, \\
& \lambda_{(1)}\bigg(\frac{\mathrm{d}\langle X_i,X_j\rangle(t)}{\mathrm{d}t}\bigg)_{1\le i,j\le n}\le L,\quad \text{a.e. } t\ge0
\end{split}
\end{equation}
hold almost surely; and 
\begin{equation}
\tau_K := \inf\big\{t \geq 0:\, X(t) \notin K\big\}.
\end{equation}
In words: $v(x)$ is the largest, across all martingale laws $\mathrm{P} \in \mathcal{P}_x$, deterministic almost sure lower bound on the exit time from $K$. The following theorem characterizes $v$, for \textit{any compact} $K\subset\mathbb{R}^n$, and is our main result.  

\begin{thm} \label{thm: vis}
Let $n\geq 2$, $k\in\{1,2,\ldots,n-1\}$, and $L\ge1$. Suppose $K\subset \mathbb{R}^n$ is compact. Then, the value function $v$ of \eqref{soc problem} is an upper semicontinuous viscosity solution of the fully nonlinear elliptic partial differential equation 
$F(\nabla v,\nabla^2 v)=1$ on $K$ with zero boundary condition (see Definition \ref{def:visc} below) where
\begin{equation} \label{def: F}
F(p, M):=\inf\bigg\{\!-\frac{1}{2}\mathrm{tr}(Ma):\, a\succeq 0,\, a p = 0,\,  \lambda_{(n-k)}(a)\geq 1,\, \lambda_{(1)}(a)\leq L\bigg\}.
\end{equation}
Suppose, in addition, that there are $T_\iota\!:\rr^n\to\rr^n$, $\iota\in(1,2]$, each given by a composition of a rotation, a dilation and a translation, and satisfying $K\subset \accentset{\circ}{T_\iota(K)}$, for which $\lim_{\iota\downarrow1} T_\iota=I$. (Here, $I$ is the identity map on $\rr^n$.) Then, the upper semicontinuous viscosity solution of $F(\nabla v,\nabla^2 v)=1$ on $K$ with zero boundary condition is unique.
\end{thm}

\begin{rmk}
\begin{enumerate}[(a)]
\item Theorem \ref{thm: vis} characterizes, in particular, the solution $T^*$ of the described relative arbitrage problem via the representation $T^*=v(U\mu(0))$. \vskip 5.5pt
\item The nonlinearity $F$ is `geometric', as defined in \cite{BSS}, i.e., for any $p \in \rr^n$, symmetric $n\times n$ matrix $M$, $c_1>0$, and $c_2\in\rr$,
\begin{equation}
F(c_1 p, c_1 M + c_2pp^\top) = c_1 F(p,M).
\end{equation}
Parabolic equations with such nonlinearities appear in weak formulations of geometric flows, and the corresponding viscosity theory was first developed in \cite{cgg,evans_motion_1991,son} for the classical mean curvature flow, and then extended in \cite{BSS,ambrosio_level_1996}. \vskip 5.5pt
\item When $L=1$, the partial differential equation $F(\nabla v,\nabla^2 v)=1$ with zero boundary condition becomes the arrival time formulation of a co-dimension mean curvature flow from \cite{ambrosio_level_1996}. For a related but different stochastic representation of these geometric flows we refer to \cite{SoTo}. \vskip 5.5pt
\item In view of the right-hand side of \eqref{soc problem}, it is natural to conjecture that the value function $v$ does not depend on $L$, at least when $K$ is convex. We were not able to show this and leave it as a tantalizing open problem. 
\end{enumerate}
\end{rmk}

The remainder of the paper is structured as follows. In Section \ref{sec2}, we show the upper semicontinuity of the value function $v$, as well as a dynamic programming principle it satisfies. In Subsections \ref{subsec: sub} and \ref{subsec super} of Section \ref{sec3}, we establish the viscosity subsolution and supersolution properties of $v$, respectively. Section \ref{sec4} is then devoted to the uniqueness of the upper semicontinuous viscosity solution under the additional assumption in Theorem~\ref{thm: vis}, finishing the proof of the latter. Finally, Section \ref{sec5} examines the continuity of the value function $v$, particularly in the case that $K$ is a polytope as in the setting of stochastic portfolio theory.   

\medskip

\noindent\textbf{Acknowledgement.} The authors would like to thank Martin Larsson and Johannes Ruf for many enlightening discussions on the subject of the paper. 

\section{Properties of the value function} \label{sec2}

The main result of this section (Proposition \ref{prop: dpp}) addresses properties of the value function $v$ from \eqref{soc problem}, including the dynamic programming principle. We start with a series of lemmas pertaining to the sets $\mathcal{P}_x$. Throughout we write $\mathbb{S}^n_+$ for the set of $n\times n$ symmetric positive semidefinite matrices. The first lemma demonstrates that the set of instantaneous covariation matrices defined by \eqref{LSS cond'} is the convex hull of the one defined by \eqref{LSS cond}. 

\begin{lemma} \label{lem: convex hull}
Let $n\ge2$ and $k\in\{1,2,\ldots,n-1\}$. Then, the convex hull of the set $\{a \in \mathbb{S}_+^n \!: \lambda_{(n-k)}(a)\geq 1\}$ is
\begin{equation}
\big\{a \in \mathbb{S}^n_+:\,\Pi_m(a)\geq m-k \text{ for }m=k+1,\, k+2,\,\ldots,\, n \big\}=:A.
\end{equation}
\end{lemma}

\smallskip

\noindent\textbf{Proof.} We first observe that the ``trace operator'' $\Pi_m$
sums the $m$ smallest eigenvalues:
$\Pi_m(a) = \sum_{i=n-m+1}^n \lambda_{(i)}(a)$, $a  \in \mathbb{S}_+^n$. Let $B$ be the convex hull of $\{a \in \mathbb{S}_+^n \!: \lambda_{(n-k)}(a) \geq 1\}$. The above formula for $\Pi_m(a)$ reveals that $\{a \in \mathbb{S}_+^n \!: \lambda_{(n-k)}(a) \geq 1\} \subseteq A$. Moreover, for any $a, \widetilde{a} \in A$ and $c \in (0,1)$,  the concavity of $\Pi_m$ implies that
\begin{equation}
	\Pi_m\big(c a + (1-c) \widetilde{a}\big) \geq c\Pi_m(a) + (1-c) \Pi_m(\widetilde{a}) \geq m-k\quad\text{for}\quad m = k+1,\, k+2,\,\ldots,\, n.
\end{equation}
Thus, $A$ is convex. Consequently, $B\subseteq A$.
	
\medskip
	
Now, suppose that $B \subsetneq A$. Choose a matrix $\overline{a} \in A$ such that $\overline{a} \notin B$. By the hyperplane separation theorem in the space of $n\times n$ symmetric matrices with inner product $\text{tr}(a\widetilde{a})$, there exists an $n\times n$ symmetric matrix $M$ and $c_1, c_2 \in \mathbb{R}$ such that
\begin{equation}
\text{tr}(\overline{a}M) < c_1 < c_2 < \text{tr}(bM),\quad 
b\in \{a \in \mathbb{S}_+^n:\, \lambda_{(n-k)}(a)\geq 1\}.
\end{equation}
Since the singleton $\{\overline{a}\}$ is compact and $\{a \in \mathbb{S}_+^n \!: \lambda_{(n-k)}(a)\geq 1\}$ is closed, the inequalities can be strict. We also observe that necessarily $M\in\mathbb{S}^n_+$ because $\text{tr}(bM)$ is lower bounded over $b \in \{a \in \mathbb{S}_+^n \!: \lambda_{(n-k)}(a)\geq 1\}$.
	
\medskip
	
Let $q_1,\,q_2,\,\ldots,\,q_n$ be orthonormal eigenvectors of $M$ with eigenvalues $\lambda_1,\,\lambda_2,\,\ldots,\,\lambda_n$ and $Q_{(1)} \ge Q_{(2)}\ge\cdots \ge Q_{(n)} \geq 0$ be the ordered sequence of $Q_i := \mathrm{tr}(\overline{a}q_i q_i^{\top})$, $i=1,\,2,\,\ldots,\,n$. The property $\overline{a}\in A$ implies that $\sum_{i=1}^m Q_{(n-i+1)} \geq m-k$ for $m = k+1,\,k+2, \ldots,\, n$. Therefore,
\begin{equation}
\begin{split}
& \, c_1 > \text{tr}(\overline{a}M) 
= \sum_{i=1}^n \lambda_i\,Q_i \\
& \geq \sum_{i=1}^n \lambda_{(i)}\,Q_{(n-i+1)} \\
&\geq \lambda_{(k+1)}\,\bigg( \sum_{i=1}^{k+1} Q_{(n-i+1)} \bigg) 
+ \sum_{i=k+2}^n \lambda_{(i)}\,Q_{(n-i+1)} \\
&= \lambda_{(k+1)} + \lambda_{(k+1)}\,\bigg( \sum_{i=1}^{k+1} Q_{(n-i+1)} - 1 \bigg) + \lambda_{(k+2)}\,Q_{(n-k-1)} + \sum_{i=k+3}^n \lambda_{(i)}\,Q_{(n-i+1)} \\
&\geq \lambda_{(k+1)} + \lambda_{(k+2)} + \lambda_{(k+2)}\,\bigg( \sum_{i=1}^{k+2} Q_{(n-i+1)} - 2 \bigg) + \sum_{i=k+3}^n \lambda_{(i)}\,Q_{(n-i+1)} \\
&\geq \cdots \geq \sum_{i=k+1}^n \lambda_{(i)} > c_2,
\end{split}
\end{equation}
which is a contradiction. It follows that $B=A$. \ep

\medskip

The next lemma yields the relative compactness of the sets $\mathcal{P}_x$.

\begin{lemma}\label{lemma: precompact}
If $S \subset \mathbb{S}^n_+$ is bounded, then the set of continuous martingale laws under which $X(0)=x\in{\mathbb R}^n$ and $\big(\frac{\mathrm{d}\langle X_i,X_j\rangle(t)}{\mathrm{d}t}\big)_{1\le i,j\le n}\in S$, \text{a.e. }$t\ge0$ almost surely is relatively compact for the topology of weak convergence. In particular, each $\mathcal{P}_x$ is relatively compact. 
\end{lemma}

\smallskip

\noindent\textbf{Proof.} Let $\mathrm{P}$ be a martingale law as described and $C<\infty$ be a constant such that 
\begin{equation}\label{C def}
\mathrm{P}\bigg(\text{tr}\bigg(\frac{\mathrm{d}\langle X_i,X_j\rangle(t)}{\mathrm{d}t}\bigg)_{1\le i,j\le n}\le C,\; \text{a.e. }t\ge0\bigg)=1.
\end{equation}
Set $\langle X \rangle=(\langle X_i,X_j\rangle)_{1\le i,j\le n}$, fix $s \geq 0$, and define 
\begin{equation}
M(t) = |X(t) - X(s)|^2 - \text{tr}\big(\langle X \rangle(t)\big) + \text{tr}\big(\langle X \rangle(s)\big), \quad t \geq s. 
\end{equation}
Since $X$ is a martingale, $M$ is a local martingale on $[s, \infty)$. Moreover, $\mathrm{P}$-almost surely, $\langle M \rangle (t) \leq 4 C \int_s^t |X(u) - X(s)|^2\,\mathrm{d}u$, $t\ge s$ by It\^o's formula and \eqref{C def}. Using the Burkholder-Davis-Gundy inequality and again \eqref{C def}, we find that for all $t\ge s$,
\begin{equation*}
\begin{split}
\mathbb{E}^{\mathrm{P}}[\langle M \rangle (t)] 
\le 4C \int_s^t \mathbb{E}^{\mathrm{P}}\big[|X(u) - X(s)|^2\big]\,\mathrm{d}u 
&\le 16 C \int_s^t \mathbb{E}^{\mathrm{P}}\big[\text{tr}\big(\langle X\rangle(u)\big) - \text{tr}\big(\langle X \rangle(s)\big)\big]\,\mathrm{d}u \\
&\leq 16C^2 \int_s^t (u-s)\,\mathrm{d}u = 8C^2(t - s)^2.
\end{split}
\end{equation*} 
Another application of the Burkholder-Davis-Gundy inequality and \eqref{C def} yields
\begin{equation}\label{moment bound}
\begin{split}
\mathbb{E}^{\mathrm{P}}\big[|X(t) - X(s)|^4\big] &= \mathbb{E}^{\mathrm{P}}\big[\big(M(t) + \text{tr}\big(\langle X \rangle(t)\big) - \text{tr}\big(\langle X \rangle(s)\big)\big)^2\big] \\
& \leq 2 \mathbb{E}^{\mathrm{P}}\big[M(t)^2\big] + 2 \mathbb{E}^{\mathrm{P}}\big[\big(\text{tr}\big(\langle X\rangle(t)\big) - \text{tr}\big(\langle X \rangle(s)\big)\big)^2\big] \\
&\leq 8 \mathbb{E}^{\mathrm{P}}[\langle M \rangle (t)] + 2C^2(t - s)^2 
\leq 66C^2(t - s)^2,\quad t\ge s.
\end{split}
\end{equation}

\medskip

In view of the bound on $\mathbb{E}^{\mathrm{P}}[|X(t) - X(s)|^4]$, the Kolmogorov continuity criterion (see, e.g., \cite[Chapter I, Theorem 2.1]{revuz_continuous_1999}) implies that for any $T\in(0,\infty)$ and $\alpha \in \big(0, \frac{1}{4}\big)$, the expectation
\[
\mathbb{E}^{\mathrm{P}} \bigg[ \bigg( \sup_{0 \leq s < t \leq T} \frac{|X(t) - X(s)|}{|t - s|^\alpha} \bigg)^4 \bigg] 
\]
is bounded uniformly over $\mathrm{P}$. This ensures the desired relative compactness, by Prokhorov's Theorem together with the relative compactness of Hölder balls in $C([0, T], \mathbb{R}^n)$ due to the Arzelà–Ascoli Theorem.  \ep

\medskip

We turn to the compactness of the sets $\mathcal{P}_x$.

\begin{lemma}\label{lemma: compact}
If $S \subset \mathbb{S}^n_+$ is a compact convex set, then the set of continuous martingale laws under which $X(0)=x\in{\mathbb R}^n$ and $\big(\frac{\mathrm{d}\langle X_i,X_j\rangle(t)}{\mathrm{d}t}\big)_{1\le i,j\le n}\in S$, $\text{a.e. }t\ge0$ almost surely is compact for the topology of weak convergence. In particular, each $\mathcal{P}_x$ is compact. 
\end{lemma}

\smallskip

\noindent\textbf{Proof.} In view of Lemma \ref{lemma: precompact}, it suffices to verify that the described set of martingale laws is closed. To this end, let $(\mathrm{P}^m)_{m=1}^\infty$ be a sequence of martingale laws as described converging weakly. Upon noting that $\langle X\rangle$ is Lipschitz under $\mathrm{P}^m$ uniformly over $m$, and thus uniformly tight, we apply Prokhorov's Theorem followed by Skorokhod's Representation Theorem to find a subsequence of $m\ge1$ and an almost sure instance $(X^m,\langle X^m\rangle)\to (X^\infty,Y^\infty)$ of the weakly convergent $\mathrm{P}^m\circ(X,\langle X\rangle)^{-1}$. Then, $X^\infty$ and $X^\infty(X^\infty)^\top-Y^\infty$ are martingales thanks to Vitali's Convergence Theorem, the moment bound \eqref{moment bound} and the uniform in $m$ Lipschitz property of $\langle X^m\rangle$. In particular, for all $0\le s<t$,
\begin{equation}
\frac{\langle X^\infty \rangle(t)-\langle X^\infty \rangle(s)}{t-s}
=\lim_{m\to\infty} \frac{\langle X^m\rangle(t)-\langle X^m\rangle(s)}{t-s}
=\lim_{m\to\infty}\,\frac{1}{t-s}\int_s^t \frac{\mathrm{d}\langle X^m\rangle(u)}{\mathrm{d}u}\,\mathrm{d}u
\end{equation}
lies in $S$  almost surely, since $S$ is convex and closed. By Lebesgue's Fundamental Theorem of Calculus and Differentiation Theorem, $\frac{\mathrm{d}\langle X^\infty \rangle(t)}{\mathrm{d}t}\in S$, $\text{a.e. }t\ge0$ almost surely. \ep

\medskip

The compactness of $\mathcal{P}_x$ lets us conclude the subsequent properties of the value function.

\begin{prop}\label{prop: dpp}
Let $n\geq 2$, $k\in\{1,2,\ldots,n-1\}$, and $L\ge1$. Suppose $K\subset \mathbb{R}^n$ is compact. Then, the value function $v$ from \eqref{soc problem} is upper semicontinuous on $\mathbb{R}^n$. Moreover, it satisfies the following dynamic programming principle: For any $x\in\mathbb{R}^n$ and any stopping time $\theta$ with respect to the filtration generated by the coordinate process $X$,
\begin{equation}\label{eq: DPP}
v(x) = \sup_{\mathrm{P}\in\mathcal{P}_x} \mathrm{P}\text{-}\mathrm{ess} \inf \, \big(\theta\wedge\tau_K + v(X(\theta))\,\mathbf{1}_{\{\theta\leq \tau_K\}}\big).
\end{equation}
In addition, the supremum in \eqref{eq: DPP} is attained by any optimizer $\mathrm{P}\in \mathcal{P}_x$ in \eqref{soc problem}.
\end{prop}

\smallskip

\noindent\textbf{Proof.} It suffices to repeat \cite[proofs of Proposition 2.2(ii), (iii)]{larsson_minimum_2022} word by word. \ep

\begin{rmk}
The above dynamic programming principle is pointwise and differs from the classical one
of stochastic optimal control which involves an expectation. This extension was first observed in \cite{ST98} for stochastic target problems and later used in \cite{soner_dynamic_2002} for geometric flows.
\end{rmk}

%
%

\section{Viscosity Solution Property of the Value Function} \label{sec3}

This section is devoted to verifying that the value function $v$ of \eqref{soc problem} is a viscosity solution to $F(\nabla v,\nabla^2 v)=1$ on $K$ with zero boundary condition, where 
\begin{equation} \label{codim k pde}
F(p, M):=\inf\bigg\{\!-\frac{1}{2}\mathrm{tr}(Ma):\, a\succeq 0,\, a p = 0,\,  \lambda_{(n-k)}(a)\geq 1,\, \lambda_{(1)}(a)\leq L\bigg\}.
\end{equation} 

\smallskip

We start with the definition of a viscosity solution in our setting, cf.~
\cite{CL, CEL, FS}. Hereby, we use upper and lower stars to denote the upper and lower semicontinuous envelopes ($\lim_{\varepsilon\downarrow0}\sup_{y:\,|y-x|<\varepsilon}$ and $\lim_{\varepsilon\downarrow0}\inf_{y:\,|y-x|<\varepsilon}$) of a function, respectively. We also write $\accentset{\circ}{K}$ for the interior of $K$. 

\begin{defn} \label{def:visc}
Let $n\geq 2$, $k\in\{1,2,\ldots,n-1\}$, $L\ge1$, and $K\subset \mathbb{R}^n$ be compact.
\begin{enumerate}[(a)]
\item A bounded function $u\!: K \to \mathbb{R}$ is a \textit{viscosity subsolution} of $F(\nabla u, \nabla^2 u)=1$ in $\accentset{\circ}{K}$ if for any $x\in\accentset{\circ}{K}$ and $\varphi\in C^2(\mathbb{R}^n)$ such that $(u^{\ast}-\varphi)(x)=\max_K (u^{\ast}-\varphi)$, it holds
\begin{equation}\label{eq: subs ineq}
F_{\ast}(\nabla \varphi(x), \nabla^2 \varphi(x))\leq 1.
\end{equation}
The function $u$ satisfies the \textit{zero boundary condition} if for any $x\in\partial K$ with $u^{\ast}(x)>0$ and $\varphi\in C^2(\mathbb{R}^n)$ such that $(u^{\ast}-\varphi)(x)=\max_K (u^{\ast}-\varphi)$, one has the inequality \eqref{eq: subs ineq}. \\
\item A bounded function $u\!: K \to \mathbb{R}$ is a \textit{viscosity supersolution} of $F(\nabla u, \nabla^2 u)=1$ in $\accentset{\circ}{K}$ if for any $x\in\accentset{\circ}{K}$ and $\varphi\in C^2(\mathbb{R}^n)$ such that $(u_{\ast}-\varphi)(x)=\min_K (u_{\ast}-\varphi)$, it holds
\begin{equation}\label{eq: supers ineq}
F^{\ast}(\nabla \varphi(x), \nabla^2 \varphi(x))\geq 1.
\end{equation}
The function $u$ satisfies the \textit{zero boundary condition} if for any $x\in\partial K$ with $u_{\ast}(x)<0$ and $\varphi\in C^2(\mathbb{R}^n)$ such that $(u_{\ast}-\varphi)(x)=\min_K (u_{\ast}-\varphi)$, one has the inequality \eqref{eq: supers ineq}. \\
\item A bounded function $u\!: K \to \mathbb{R}$ is a \textit{viscosity solution} of $F(\nabla u, \nabla^2 u)=1$ on $K$ with \textit{zero boundary condition} if it is a viscosity subsolution in $\accentset{\circ}{K}$ satisfying the zero boundary condition and a viscosity supersolution in $\accentset{\circ}{K}$ satisfying the zero boundary condition. 
\end{enumerate}
\end{defn}

\smallskip

In view of Definition \ref{def:visc}, our first aim is to find $F_*(p,M)$ and $F^*(p,M)$. To this end, let
\begin{equation}
M_p:=\begin{cases}
\big(I - pp^{\top}/|p|^2\big)M\big(I - pp^{\top}/|p|^2\big)+\min\big(\lambda_{(n)}(M),0\big)\,pp^{\top}/|p|^2,
\quad\text{if }\,p\neq0, \\
M,\quad\text{if  }\,p=0,
\end{cases}
\end{equation}
where $I$ is the $n\times n$ identity matrix. Observing that $\mathrm{tr}(Ma)=\mathrm{tr}(M_pa)$ for all $a\succeq 0$ with $ap=0$ in the definition \eqref{codim k pde} of $F$, and writing the symmetric $M_p$ as a linear combination of outer products, we see that
\begin{equation}\label{F spec}
\begin{split}
F(p, M)=&-\frac{1}{2}\sum_{i=1}^{n-k} \big(L\lambda_{(i)}(M_p)\,\mathbf{1}_{\{\lambda_{(i)}(M_p)>0\}}
+\lambda_{(i)}(M_p)\,\mathbf{1}_{\{\lambda_{(i)}(M_p)\le 0\}}\big) \\
&-\frac{1}{2}\sum_{i=n-k+1}^n L\lambda_{(i)}(M_p)\,\mathbf{1}_{\{\lambda_{(i)}(M_p)>0\}}.
\end{split}
\end{equation}
We are now ready to compute $F_*$ and $F^*$. 

\begin{lemma} \label{compute F}
The nonlinearity $F$ satisfies $F_*=F^*=F$ on $({\mathbb R}^n\backslash\{0\})\times\mathbb{S}^n$, and $F_*=F$ on $\{0\}\times\mathbb{S}^n$. Moreover, for all $M\in\mathbb{S}^n$,
\begin{equation}
\begin{split}
F^*(0,M) =&-\frac{1}{2}\sum_{i=2}^{n-k+1} \big(L\lambda_{(i)}(M)\,\mathbf{1}_{\{\lambda_{(i)}(M)>0\}}
+\lambda_{(i)}(M)\,\mathbf{1}_{\{\lambda_{(i)}(M)\le 0\}}\big) \\
&-\frac{1}{2}\sum_{i=n-k+2}^n L\lambda_{(i)}(M)\,\mathbf{1}_{\{\lambda_{(i)}(M)>0\}}.
\end{split}
\end{equation}
\end{lemma}

\smallskip

\noindent\textbf{Proof.} The nonlinearity $F$ is continuous on  $({\mathbb R}^n\backslash\{0\})\times\mathbb{S}^n$ thanks to the continuity of $(p,M)\mapsto M_p$, $M_p\mapsto(\lambda_{(1)}(M_p),\lambda_{(2)}(M_p),\ldots,\lambda_{(n)}(M_p))$, $\lambda\mapsto L\lambda\,\mathbf{1}_{\{\lambda>0\}}+\lambda\,\mathbf{1}_{\{\lambda\le0\}}$ and $\lambda\mapsto L\lambda\,\mathbf{1}_{\{\lambda>0\}}$. Hence, $F_*=F^*=F$ on $({\mathbb R}^n\backslash\{0\})\times\mathbb{S}^n$. 

\medskip

%

Next, given any $M\in\mathbb{S}^n$, we pick a sequence $(p^m,M^m)_{m=1}^\infty$ in ${\mathbb R}^n\times\mathbb{S}^n$ converging to $(0,M)$ and such that $\lim_{m\to\infty} F(p^m,M^m)=F_*(0,M)$. Then, by the definition of $F$ in \eqref{codim k pde} and the formula \eqref{F spec} together with the continuity of $M\mapsto(\lambda_{(1)}(M),\lambda_{(2)}(M),\ldots,\lambda_{(n)}(M))$, $\lambda\mapsto L\lambda\,\mathbf{1}_{\{\lambda>0\}}+\lambda\,\mathbf{1}_{\{\lambda\le0\}}$ and $\lambda\mapsto L\lambda\,\mathbf{1}_{\{\lambda>0\}}$,
\begin{equation}
F_*(0,M)=\lim_{m\to\infty} F(p^m,M^m)\geq \limsup_{m\to\infty} F(0,M^m) =F(0, M).
\end{equation}
Consequently, $F_*(0,M)=F(0,M)$. 

\medskip

Finally, we compute the upper semicontinuous envelope $F^*$ at $(0,M)$.
Let $(p^m,M^m)_{m=1}^\infty$ be a sequence in ${\mathbb R}^n\times\mathbb{S}^n$ converging to $(0,M)$ with $\lim_{m\to\infty} F(p^m,M^m)=F^*(0,M)$. By the Poincar\'{e} Separation Theorem, $\lambda_{(i)}(M^m_{p^m})\ge\lambda_{(i+1)}(M^m)$, $i\in\{1,2,\ldots,n-1\}$. Since the functions $\lambda\mapsto L\lambda\,\mathbf{1}_{\{\lambda>0\}}+\lambda\,\mathbf{1}_{\{\lambda\le0\}}$ and $\lambda\mapsto L\lambda\,\mathbf{1}_{\{\lambda>0\}}$ are non-decreasing, applying the formula \eqref{F spec} to $F(p^m,M^m)$ we obtain
\begin{equation}
\begin{split}
F^*(0,M)&=\lim_{m\to\infty} F(p^m,M^m) \\
&\le \liminf_{m\to\infty}
\bigg(\!-\frac{1}{2}\sum_{i=2}^{n-k+1} \big(L\lambda_{(i)}(M^m)\,\mathbf{1}_{\{\lambda_{(i)}(M^m)>0\}}
+\lambda_{(i)}(M^m)\,\mathbf{1}_{\{\lambda_{(i)}(M^m)\le 0\}}\big) \\
&\qquad\qquad\;\;\; -\frac{1}{2}\sum_{i=n-k+2}^n L\lambda_{(i)}(M^m)\,\mathbf{1}_{\{\lambda_{(i)}(M^m)>0\}}\bigg) \\
&=-\frac{1}{2}\sum_{i=2}^{n-k+1} \big(L\lambda_{(i)}(M)\,\mathbf{1}_{\{\lambda_{(i)}(M)>0\}}
+\lambda_{(i)}(M)\,\mathbf{1}_{\{\lambda_{(i)}(M)\le 0\}}\big) \\
&\quad -\frac{1}{2}\sum_{i=n-k+2}^n L\lambda_{(i)}(M)\,\mathbf{1}_{\{\lambda_{(i)}(M)>0\}},
\end{split}
\end{equation} 
where the last equality is due to the continuity of $M\mapsto(\lambda_{(1)}(M),\lambda_{(2)}(M),\ldots,\lambda_{(n)}(M))$, $\lambda\mapsto L\lambda\,\mathbf{1}_{\{\lambda>0\}}+\lambda\,\mathbf{1}_{\{\lambda\le0\}}$ and $\lambda\mapsto L\lambda\,\mathbf{1}_{\{\lambda>0\}}$.

\medskip

To show the reversed inequality, we pick the sequence $\big(\frac{q_1}{m}, M\big)_{m=1}^\infty$ converging to $(0,M)$, where $q_1$ is an eigenvector of $M$ with the eigenvalue $\lambda_{(1)}(M)$. Then, writing $M$ as a linear combination of outer products in the definition \eqref{codim k pde} of $F\big(\frac{q_1}{m}, M\big)$, we infer 
\[
\begin{split}
F^*(0, M) \ge \limsup_{m\to\infty} F\Big(\frac{q_1}{m}, M\Big) 
= &-\frac{1}{2}\sum_{i=2}^{n-k+1} \big(L\lambda_{(i)}(M)\,\mathbf{1}_{\{\lambda_{(i)}(M)>0\}}
+\lambda_{(i)}(M)\,\mathbf{1}_{\{\lambda_{(i)}(M)\le 0\}}\big) \\ 
&-\frac{1}{2}\sum_{i=n-k+2}^n L\lambda_{(i)}(M)\,\mathbf{1}_{\{\lambda_{(i)}(M)>0\}}.
\end{split}
\]
This concludes the proof of the lemma. \ep

\medskip

As a further preparation, we consider the following example.

\begin{exm} \label{ex: ball}
Let $K:=\overline{B_r(0)}$, the closed ball of radius $r>0$ around the origin in $\mathbb{R}^n$. Then, the value function $v$ of \eqref{soc problem} is given by
\begin{equation}\label{eq: radial v}
v(x) = \frac{\max(r^2 - |x|^2,0)}{n-k},\quad x\in \mathbb{R}^n. 
\end{equation}
\end{exm}

\begin{proof}
For any $x\in K$ and $\mathrm{P}\in\mathcal{P}_x$, Itô's formula and the first inequality in \eqref{eq: constraint} with $m=n$ yield
\[
|X(t)|^2 = |x|^2 + 2\int_0^t X(s)^{\top}\,\mathrm{d}X(s) + \text{tr}\big(\langle X\rangle(t)\big) 
\geq |x|^2 + 2\int_0^t X(s)^{\top}\,\mathrm{d}X(s) + (n-k)t.
\]
Upon evaluating at $\tau_K\wedge t$ and taking the expectation, we deduce
\begin{equation}
\mathrm{P}\text{-ess} \inf \tau_K
\le \lim_{t\uparrow\infty} \mathbb{E}[\tau_K\wedge t] 
\le \liminf_{t\uparrow\infty} \frac{\mathbb{E}[|X(\tau_K\wedge t)|^2] - |x|^2}{n-k} 
\le \frac{r^2 - |x|^2}{n-k}.
\end{equation}

\smallskip

For the reversed inequality, let $x\in K\backslash\{0\}$ and $n' := n-k+1$. Since the coordinates of $x$ can be relabeled, we may assume that the first $n'$ coordinates of $x$, denoted by $x_{[n']}$, satisfy $x_{[n']}\neq0$. Consider $\mathrm{P}^{\ast}\in\mathcal{P}_x$ under which the first $n'$ coordinates of $X$ follow
\begin{equation}\label{n' SDE}
\mathrm{d}X_{[n']}(t) = a(X_{[n']}(t))^{1/2}\,\mathrm{d}{W(t)},
\end{equation}
where $a(y) := I - yy^{\top}/|y|^2$ when $y\neq 0$, $a(0):=I$, and $W$ is an $n'$-dimensional standard Brownian motion. The remaining coordinates of $X$ are chosen to be constant. Then, $a^{1/2}$ is continuous on ${\mathbb R}^{n'} \setminus \{0\}$, and $a(y)^{1/2}y = 0$. Thus, for any $\varepsilon\in(0,|x_{[n']}|)$, by Itô's formula,
\[
|X_{[n']}(t)|^2 = |x_{[n']}|^2 + (n-k)t,\quad t\le\inf\big\{t'\ge0:\,|X_{[n']}(t')|\le\varepsilon\big\}.
\]
Consequently, \eqref{n' SDE} has a global weak solution satisfying $|X_{[n']}(t)|\ge|x_{[n']}|$, $t\ge0$. In addition, $\tau_K = (r^2 - |x|^2) / (n-k)$ almost surely under $\mathrm{P}^{\ast}$.

\medskip
%

For $x=0$, consider a sequence $(x^m)_{m=1}^\infty$ in $K\backslash\{0\}$ going to $x$. By the compactness of $\mathcal{P}_0$ (Lemma \ref{lemma: compact}), the associated sequence $(\mathrm{P}^{\ast}_{x^m})_{m=1}^\infty$ has a subsequence going to a $\mathrm{P}^{\ast}\in\mathcal{P}_0$. Then, it holds $|X(t)|^2 = (n-k) t$, $t \geq 0$, thus $\tau_K = r^2 / (n-k)$, almost surely under $\mathrm{P}^{\ast}$.
\end{proof}

\subsection{Subsolution property of the value function} \label{subsec: sub}

We are now ready to verify that the value function $v$ of \eqref{soc problem} satisfies Definition \ref{def:visc}(a). Since $v^*=v$ (Proposition \ref{prop: dpp}) and $F_*=F$ (Lemma \ref{compute F}), we may replace $v^*$ by $v$ and $F_*$ by $F$ in Definition \ref{def:visc}(a). Moreover, $v(x)>0$ for all $x\in \accentset{\circ}{K}$ by Example \ref{ex: ball}. Thus, we only need to study $x\in K$ with $v(x)>0$. Let $\varphi\in C^2(\mathbb{R}^n)$ be a test function such that $\varphi\ge v$ on $K$ and $\varphi(x)=v(x)$. Since $\varphi$ can be replaced by $\varphi+|\cdot-x|^4$, we may assume that $\varphi>v$ on $K\backslash\{x\}$.

\medskip

We prove the inequality \eqref{eq: subs ineq} by contradiction. Suppose that $F(\nabla\varphi(x),\nabla^2\varphi(x))>1$. Then, we let 
\begin{equation}
\mathcal{A}:=\big\{a\succeq0:\,\lambda_{(n-k)}(a)\ge1,\,\lambda_{(1)}(a)\le L\big\}
\end{equation}
and claim the existence of an $\varepsilon \in \big(0, \sqrt{v(x)(n-k)}\big)$ such that for all $a \in \mathcal{A}$ and all $y \in B_{\varepsilon}(x)\cap K$, the following implication holds:
\begin{equation}\label{eq: subsol ineq}
1 + \frac{1}{2} \text{tr}\big(a\nabla^2\varphi(y)\big) > 0 \;\;\implies\;\; \nabla \varphi(y)^{\top} a \nabla \varphi(y)\geq \varepsilon. 
\end{equation} 
Indeed, otherwise there would exist a sequence $(a^m,x^m)_{m=1}^\infty$ in  $\mathcal{A}\times K$ with $x^m \to x$,
\begin{equation}
1+\frac{1}{2}\mathrm{tr}\big(a^m \nabla^2\varphi(x^m)\big) > 0,\;\; m\in\mathbb{N},\quad
\text{and}\;\; \nabla\varphi(x^m)^{\top}a^m\,\nabla\varphi(x^m) \to 0.
\end{equation}
Since $\mathcal{A}$ is compact, $(a^m)_{m=1}^\infty$ would admit a subsequence converging to some $a\in\mathcal{A}$. Taking $m\to\infty$ in accordance with that subsequence, we would arrive at
\begin{equation}\label{subsol ineq contr}
1+\frac{1}{2}\mathrm{tr}\big(a\nabla^2\varphi(x)\big) \geq 0\quad\text{and}\quad \nabla\varphi(x)^{\top}a\nabla\varphi(x) = 0.
\end{equation}
The latter equation implies $a^{1/2}\nabla\varphi(x)=0$, thus $a\nabla\varphi(x)=0$. Therefore, \eqref{subsol ineq contr} contradicts $F(\nabla\varphi(x),\nabla^2\varphi(x))>1$ (recall the definition \eqref{codim k pde} of $F$).

\medskip

Next, we fix an optimizer $\mathrm{P}\in\mathcal{P}_{x}$ on the right-hand side of \eqref{soc problem} and let 
\begin{equation}
\theta := \inf\big\{t \geq 0:\, X(t) \notin B_{\varepsilon}(x)\big\} \wedge v(x). 
\end{equation}
Thanks to Example \ref{ex: ball}, $\tau_K\ge v(x)>\varepsilon^2/(n-k)\ge \mathrm{P}\text{-ess} \inf \tau_{\,\overline{B_{\varepsilon}(x)}}$. Hence, $\partial B_{\varepsilon}(x)\cap K \neq \varnothing$ and $\mathrm{P}(X({\theta}) \in \partial B_{\varepsilon}(x)\cap K)>0$. Since $\theta\leq \tau_K$ under $\mathrm{P}$, the dynamic programming principle (Proposition \ref{prop: dpp}) together with $\delta := \min_{\partial B_{\varepsilon}(x)\cap K} (\varphi-v)>0$ imply
\begin{equation}\label{delta gap}
\varphi(x) = v(x) \leq t\wedge \theta + v(X(t\wedge \theta)) \leq t\wedge \theta + \varphi(X(t \wedge \theta))- \delta\,\mathbf{1}_{[\theta, \infty)}(t)\,\mathbf{1}_{\{X(\theta)\in\partial B_{\varepsilon}(x)\cap K\}}.
\end{equation}

\smallskip

We proceed by writing $a(t)$, $\alpha(t)$, and $\mathcal{S}$ for $\frac{\mathrm{d}\langle X\rangle(t)}{\mathrm{d}t}$, $1+\frac{1}{2}\text{tr}\big(a(t)\nabla^2\varphi(X(t))\big)$, and
$$ 
\Big\{ s \in [0, \theta) : 1 + \frac{1}{2} \text{tr}\big(a(s) \nabla^2 \varphi(X(s))\big) > 0 \Big\},
$$
respectively. Starting from \eqref{delta gap}, applying It\^o's formula to $\varphi(X(t\wedge\theta))$, introducing the auxiliary process 
$$\widetilde{X}(t) = X(t) + \varepsilon^{-1}  \int_{0}^{t} \alpha(s)\, a(s)\, \nabla\varphi(X(s)) \,\mathbf{1}_{\mathcal S}(s) \,\mathrm{d}s,\quad t\ge0,
$$
and using \eqref{eq: subsol ineq}, we deduce
\begin{align*}
&\,\delta\,\mathbf{1}_{[\theta, \infty)}(t)\,\mathbf{1}_{\{X(\theta) \in \partial B_{\varepsilon}(x)\cap K\}} \\
& \leq t\wedge \theta + \varphi(X(t\wedge \theta)) - \varphi(x) \\
& = \int_0^{t \wedge \theta} \nabla\varphi(X(s))^{\top}\,\mathrm{d}X(s) 
+ \int_0^{t \wedge \theta} 
1 + \frac{1}{2}\text{tr}\big(a(s)\nabla^2\varphi(X(s))\big)\,\mathrm{d}{s} \\
&\leq \int_0^{t \wedge \theta} \nabla\varphi(X(s))^{\top}\,\mathrm{d}X(s) 
+ \int_0^{t \wedge \theta}  \alpha(s)\,\mathbf{1}_{\mathcal S}(s)\,\mathrm{d}{s} \\
&= \int_0^{t \wedge \theta} \nabla\varphi(X(s))^{\top}\,\mathrm{d}\widetilde{X}(s) 
+ \int_0^{t \wedge \theta} \alpha(s)\big(1 - \varepsilon^{-1} \nabla\varphi(X(s))^{\top}a(s)\nabla\varphi(X(s))\big)\,\mathbf{1}_{\mathcal S}(s)\,\mathrm{d}s \\
&\le \int_0^{t \wedge \theta} \nabla\varphi(X(s))^{\top}\,\mathrm{d}\widetilde{X}(s).
\end{align*}

\smallskip
    
Finally, consider the exponential local martingale $Z$ given by
\begin{equation}
\frac{\mathrm{d}Z(t)}{Z(t)} =-\varepsilon^{-1} \alpha(t)\,\mathbf{1}_{\mathcal S}(t)\, \nabla\varphi(X(t))^{\top}\,\mathrm{d}{X(t)}, \quad Z_0 = 1.
\end{equation}
Due to the boundedness of $a(\cdot)$ and the boundedness of $\nabla^2 \varphi$, $\nabla\varphi$ on $B_\varepsilon(x)$, the process $Z$ is well-defined. It\^o's formula shows that $Z(\cdot)\int_{0}^{\cdot \wedge \theta} \nabla\varphi(X(s))^{\top} \,\mathrm{d}\widetilde{X}(s)$ is a nonnegative local martingale, hence a supermartingale. Moreover, $\theta\le v(x)$ renders the Optional Sampling Theorem applicable and we find via the final display of the previous paragraph:
\begin{equation}
0 < \delta \mathbb{E}\left[Z(\theta)\,\mathbf{1}_{\{X(\theta) \in \partial B_{\varepsilon}(x)\cap K\}} \right] \leq \mathbb{E}\left[ Z(\theta) \int_{0}^{\theta} \nabla\varphi(X(s))^{\top} \,\mathrm{d}\widetilde{X}(s) \right] \leq 0, 
\end{equation}
a contradiction. The proof of the subsolution property is complete. \ep

\subsection{Supersolution property of the value function} \label{subsec super}

Since $v\ge 0$, it suffices to check the supersolution inequality \eqref{eq: supers ineq} for $x\in\accentset{\circ}{K}$. Fix any $x\in\accentset{\circ}{K}$, and let $\varphi\in C^2(\mathbb{R}^n)$ satisfy $\varphi \le v_*$ on $K$ and $\varphi(x)=v_*(x)$. Since we can study $\varphi-\varepsilon|\cdot-x|^2$ and then pass to the limit $\varepsilon\downarrow0$, we may assume  $\varphi<v_*$ on $K\backslash\{x\}$ and that $\nabla^2 \varphi(x)$ is non-singular. We distinguish two cases: $\nabla\varphi(x)\neq0$ and $\nabla\varphi(x)=0$.

\medskip

\noindent\textbf{Case 1: $\nabla\varphi(x)\neq 0$.} In this case, $F^*(\nabla\varphi(x),\nabla^2\varphi(x))=F(\nabla\varphi(x),\nabla^2\varphi(x))$. We argue by contradiction and suppose that $F(\nabla \varphi(x),\nabla^2\varphi(x))<1$. Then, there exists an $a\in{\mathcal A}$ with 
\begin{equation}\label{case 1 a cond}
1+\frac{1}{2}\text{tr}\big(a\nabla^2\varphi(x)\big)>0\quad\text{and}\quad a\nabla\varphi(x) = 0.
\end{equation}
By the Spectral Theorem, $a=\sum_{i=1}^n \lambda_{(i)}(a)\,q_i q_i^\top$ where 
$q_1,\,q_2,\,\ldots,\,q_n$ are orthonormal eigenvectors of $a$. We can modify $a$ such that $\lambda_{(1)}(a),\,\lambda_{(2)}(a),\,\ldots,\,\lambda_{(n-k)}(a)\in(1,L)$ and \eqref{case 1 a cond} remains true. In view of $a\nabla\varphi(x)=0$, it holds $\lambda_{(i)}(a)\, q_i^{\top}\nabla\varphi(x)=0$ for all $i$. In particular, $q_i^{\top} \nabla \varphi(x) = 0$ for $i=1,\,2,\,\ldots,\, n-k$.

\medskip

Next, we introduce the $n\times n$ matrices
\begin{equation}
S_i = \frac{\lambda_{(i)}(a)^{1/2}}{|\nabla\varphi(x)|^2}\,
\big(q_i\nabla\varphi(x)^{\top}-\nabla\varphi(x)q_i^{\top}\big),\quad i=1,\,2,\,\ldots,\,n.
\end{equation}
Observe that $S_i\nabla \varphi(x) = \lambda_{(i)}(a)^{1/2}\,q_i$, $i=1,\,2,\,\ldots,\,n$. Now, let $\Sigma\!:{\mathbb R}^n\to{\mathbb R}^{n\times n}$ be such that the $i$-th column of each $\Sigma(y)$ is $S_i\nabla\varphi(y)$ for $i=1,\,2,\,\ldots,\,n$. Then, 
\begin{equation}
\Sigma(x)\,\Sigma^{\top}(x)=a \quad\text{and}\quad 1+\frac{1}{2}\text{tr}\big(\Sigma(x)\,\Sigma^{\top}(x)\,\nabla^2\varphi(x)\big) 
= 1 + \frac{1}{2}\text{tr}\big(a\nabla^2\varphi(x)\big)>0. 
\end{equation}
By the continuity of $\nabla\varphi$, $M\mapsto(\lambda_{(1)}(M),\lambda_{(2)}(M),\ldots,\lambda_{(n)}(M))$ and $\nabla^2\varphi$, there exists an $\varepsilon>0$ with the properties  $\overline{B_{\varepsilon}( x)}\subset\accentset{\circ}{K}$ and that for all $y\in B_{\varepsilon}(x)$,
\begin{equation}\label{Sigma cond}
\lambda_{(n-k)}\big(\Sigma(y)\,\Sigma^{\top}(y)\big)\ge 1,\quad
\lambda_{(1)}\big(\Sigma(y)\,\Sigma^{\top}(y)\big)\le L\quad \text{and} \quad 1+\frac{1}{2}\text{tr}\big(\Sigma(y)\,\Sigma^{\top}(y)\,\nabla^2\varphi(y)\big) \ge 0.
\end{equation}

\smallskip

Further, for $y\in B_{\varepsilon}(x)$, consider $\mathrm{P}_y \in \mathcal{P}_y$ under which the coordinate process $X$ follows the stochastic differential equation
 \begin{equation}
 \mathrm{d}{X(t)} =\sum_{i=1}^{n} \big(\mathbf{1}_{[0,\tau_{B_\varepsilon(x)})}(t)\, S_i\nabla\varphi(X(t)) + \mathbf{1}_{[\tau_{B_\varepsilon(x)}, \infty)}(t)\, e_i\big)\,\mathrm{d}W_i(t)
\end{equation}
where $e_1,\, e_2,\,\ldots,\, e_n$ is the standard basis of ${\mathbb R}^n$ and $W_1,\,W_2,\,\ldots,\,W_n$ are independent one-dimensional standard Brownian motions. Since $\nabla\varphi$ is continuous and $\frac{\mathrm{d}\langle X\rangle(t)}{\mathrm{d}t}\!=\!(\Sigma\Sigma^{\top})(X(t))$, $t\in[0,\tau_{B_\varepsilon(x)})$, each $\mathrm{P}_y$ is a well-defined element of $\mathcal{P}_y$. By Example \ref{ex: ball}, $\mathbb{E}^y[\tau_{B_\varepsilon(x)}]<\infty$ and we may apply It\^o's formula to $\varphi(X(\tau_{B_\varepsilon(x)}))$:
\begin{equation}\label{essinf LBD}
\begin{split}
\tau_{B_\varepsilon(x)} + \varphi(X(\tau_{B_\varepsilon(x)})) & = \varphi(y) + \sum_{i=1}^n \int_0^{\tau_{B_\varepsilon(x)}}  \nabla\varphi(X(t))^{\top} S_i \nabla\varphi(X(t))\,\mathrm{d} W_i(t) \\
& \quad +\int_0^{\tau_{B_\varepsilon(x)}} 1 + \frac{1}{2} \text{tr}\big(\Sigma(X(t))\,\Sigma^{\top}(X(t))\,\nabla^2\varphi(X(t))\big)\,\mathrm{d}t \\ 
& \ge \varphi(y),
\end{split}
\end{equation}
where we have used the skew-symmetry of the $S_i$'s to conclude that the martingale term vanishes, as well as \eqref{Sigma cond}.

\medskip

Lastly, with $\delta:=\min_{\partial B_\varepsilon(x)} (v-\varphi) > 0$, we see from the dynamic programming principle (Proposition \ref{prop: dpp}) and \eqref{essinf LBD} that
\begin{equation*}
v(y) \geq \mathrm{P}_y \text{-}\mathrm{ess} \inf \,\big(\tau_{B_\varepsilon(x)} + v(X(\tau_{B_\varepsilon(x)}))\big)
\ge \mathrm{P}_y \text{-}\mathrm{ess} \inf \,\big(\tau_{B_\varepsilon(x)} + \varphi(X(\tau_{B_\varepsilon(x)}))\big)+ \delta 
\ge \varphi(y) + \delta .
\end{equation*}
Taking the limit $y\to x$ along a sequence satisfying $v(y)\to v_*(x)$, we end up with 
\begin{equation}
\varphi(x) = v_*(x)\geq  \varphi(x) + \delta, 
\end{equation}
which is the desired contradiction. 

\medskip

\noindent\textbf{Case 2: $\nabla\varphi(x)=0$.} We aim for a reduction to Case 1. For simplicity of notation, we assume that $x=0$ and that $\nabla^2\varphi(0)$ is a diagonal matrix, which can be achieved by a translation followed by a rotation. For starters, we construct a sequence $(\varphi^m)_{m=1}^\infty$ in $C^2(\rr^n)$ such that
\begin{enumerate}[(a)]
\item $\varphi^m(0)=v_*(0)$ and $\varphi^m<v_*$ on $K\backslash\{0\}$, for all $m\ge1$; 
\item $\varphi^m(y)=v_*(0)-\frac{1}{2}y^\top M^my$, $y\in\overline{B_{\varepsilon^m}(0)}$ with a non-singular diagonal $M^m$ and an $\varepsilon^m>0$, for all $m\ge1$; 
\item and $\lim_{m\to\infty} M^m=\nabla^2\varphi(0)$. 
\end{enumerate}
To this end, for $m\ge1$, we let $M^m:=\nabla^2\varphi(0)-\frac{\varepsilon^0}{m}\,I$ and choose $\varepsilon^0,\varepsilon^m>0$ small enough to ensure the non-singularity of $M^m$ and $\varphi(0)+\frac{1}{2}y^\top M^m y\le \varphi(y)$, $y\in \overline{B_{2\varepsilon^m}(0)}\subset K$. Subsequently, we pick $\varphi^m\in C^2(\rr^n)$ satisfying $\varphi^m(y)\le\varphi(0)+\frac{1}{2}y^\top M^m y$, $y\in \overline{B_{2\varepsilon^m}(0)}$ with $\varphi^m(y)=\varphi(0)+\frac{1}{2}y^\top M^m y$, $y\in \overline{B_{\varepsilon^m}(0)}$, as well as $\varphi^m(y)<\min_K v_*$ on $\rr^n\backslash \overline{B_{2\varepsilon^m}(0)}$. Then, $\varphi^m(0)=\varphi(0)=v_*(0)$, (b), and (c) hold by construction. Moreover, for $m\ge1$,
\begin{eqnarray}
&& v_*(y)-\varphi^m(y)\ge v_*(y)-\varphi(y)>0,\quad y\in \overline{B_{2\varepsilon^m}(0)}\backslash\{0\}, \\
&& v_*(y)-\varphi^m(y)>v_*(y)-\min_K v_*\ge 0,\quad y\in K\backslash \overline{B_{2\varepsilon^m}(0)},
\end{eqnarray}
 so that (a) also holds. 
 
\medskip

We proceed to the main line of reasoning. For the desired $F^*(0,\nabla^2\varphi(0))\ge1$, it suffices to verify $F^*(0,M^m)\ge 1$ for each $m\ge1$ thanks to property (c). Thus, we fix an $m\ge1$ and consider auxiliary $\varphi^m(\,\cdot\,;\eta):\,\rr^n\to\rr$, $y\mapsto\varphi^m(y)-y^\top\eta$ for $\eta\in B_1(0)$. First, suppose that there exists a sequence $(\eta^\ell)_{\ell=1}^\infty$ such that $\lim_{\ell\to\infty} |\eta^\ell|=0$ and $\nabla\varphi^m(y^\ell;\eta^\ell)\neq0$, $\ell\ge1$ where each $y^\ell$ is a minimizer of $v_*-\varphi^m(\,\cdot\,;\eta^\ell)$ over $\overline{B_{\varepsilon^m}(0)}$. Then, $y^\ell\in B_{\varepsilon^m}(0)$ for all $\ell\ge1$ large enough, and arguing as in Case 1 we would obtain $F^*(\nabla\varphi^m(y^\ell;\eta^\ell),M^m)\ge1$ for those $\ell$. Due to property (a), $\lim_{\ell\to\infty} y^\ell=0$, and therefore 
\begin{equation}
\lim_{\ell\to\infty} \nabla\varphi^m(y^\ell;\eta^\ell)
=\lim_{\ell\to\infty} \big(\nabla\varphi^m(y^\ell)-\eta^\ell\big)
=\lim_{\ell\to\infty} (M^m y^\ell-\eta^\ell)=0.
\end{equation}
Hence, $F^*(0,M^m)\ge1$, as desired.

\medskip

If a sequence $(\eta^\ell)_{\ell=1}^\infty$ as above does not exist, there is an $\overline{\eta}>0$ such that for all $\eta\in B_{\overline{\eta}}(0)$ and all minimizers $y^\eta$ of $v_*-\varphi^m(\,\cdot\,;\eta)$ over $\overline{B_{\varepsilon^m}(0)}$, it holds $\nabla\varphi^m(y^\eta;\eta)=0$. We note that $\nabla\varphi^m(y^\eta;\eta)=0$ amounts to $M^m y^\eta=\eta$. Recalling that $M^m$ is diagonal and non-singular, we conclude that $\{y^\eta:\,\eta\in B_{\overline{\eta}}(0)\}$ contains an open ball $B\subset \overline{B_{\varepsilon^m}(0)}$ around $0$. For all $y^\eta\in B$, we have
\begin{equation}
v_*(y^\eta)-\varphi^m(y^\eta;\eta)=\min_{\overline{B_{\varepsilon^m}(0)}}\, (v_*-\varphi^m(\,\cdot\,;\eta)),\quad
\nabla\varphi^m(y^\eta;\eta)=0,\quad
\nabla^2\varphi^m(y^\eta;\eta)=M^m.
\end{equation}
Hence, there exists a constant $C<\infty$ such that $|v_*(\widetilde{y})-v_*(y)|\le C|\widetilde{y}-y|^2$, $y,\widetilde{y}\in B$. Consequently, $v_*\equiv c$ on $B$ for some $c\in\rr$. But then, combining the dynamic programming principle (Proposition \ref{prop: dpp}) and Example \ref{ex: ball}, we find a $\delta>0$ such that $v(y)\ge\delta+c$, $y\in\frac{1}{2}B$. This yields $c=v_*(y)\ge\delta+c$, $y\in\frac{1}{2}B$, a contradiction, ruling out the scenario under consideration. The proof of the supersolution property is finished. \ep

\section{Uniqueness} \label{sec4}

In this section, we show the next proposition, which completes the proof of Theorem \ref{thm: vis}.

\begin{prop} \label{prop: uniq}
In the setting of Theorem \ref{thm: vis}, suppose that there are $T_\iota\!:\rr^n\to\rr^n$, $\iota\in(1,2]$, each given by a composition of a rotation, a dilation and a translation, and satisfying $K\subset \accentset{\circ}{T_\iota(K)}$, for which $\lim_{\iota\downarrow1} T_\iota=I$. Then, the upper semicontinuous viscosity solution of $F(\nabla v,\nabla^2v)=1$ on $K$ with zero boundary condition is unique.  	
\end{prop}

\begin{rmk}
Mean curvature flows of any dimension are invariant under rotations, dilations, and translations (used in our assumption). This property is exploited in \cite{BSS,SS} to prove several statements about the weak flows. Also, a similar condition without a rotation is used in \cite{larsson_minimum_2022}. Our assumption is satisfied, for example, by all compact convex $K\subset\rr^n$ with nonempty interior.

\end{rmk}

The proof of Proposition \ref{prop: uniq} relies on the following two theorems of independent interest. 

\begin{thm}[Maximum Principle] \label{thm: maximum}
In the general setting of Theorem \ref{thm: vis}:
\begin{enumerate}[(a)]
\item If $u$ is an upper semicontinuous viscosity subsolution of $F(\nabla u,\nabla^2u)=1$ on $K$ and $w$ is a lower semicontinuous viscosity supersolution of $F(\nabla w,\nabla^2w)=1$ on $K$, then there exists a point $x \in \partial K$ at which the difference $u-w$ achieves its maximum over $K$. \\
\item If, in addition, $u$ satisfies the zero boundary condition, and $w$ is a lower semicontinuous viscosity supersolution of $F(\nabla w,\nabla^2w)=1$ on some compact $K'\subset\rr^n$ satisfying the zero boundary condition, where $K\subset\accentset{\circ}{K'}$, then $u \le w$ on $K$.
\end{enumerate}
\end{thm} 

\noindent\textbf{Proof.} We show both conclusions in parallel. Therein, we may replace $u$ by $u^\kappa:=\kappa u$ where $\kappa\in(0,1)$, since both (a) and (b) can be obtained by passing to the limit $\kappa\uparrow1$ at the end. (In the case of (a), any subsequential limit of $x^\kappa\in\partial K$ achieves the maximum of $u-w$ over $K$ thanks to the upper semicontinuity of $u-w$.) Now, we fix a $\kappa\in(0,1)$, and for $\varepsilon>0$ consider the functions
\begin{equation}
\Phi^\varepsilon(x,y) := u^\kappa(x) - w(y) - \varepsilon^{-1} |x-y|^4
\end{equation}
on $K\times K$ in the case of (a) and on $K\times K'$ in the case of (b). Let $(x^\varepsilon,y^\varepsilon)$ be a maximizer of $\Phi^\varepsilon$. By compactness, there is a sequence of $(x^\varepsilon,y^\varepsilon)$ converging to a limit $(x^0,y^0)$ along a sequence of $\varepsilon\downarrow0$. The inequality $\Phi^\varepsilon(x^\varepsilon,y^\varepsilon)\ge u^\kappa(x)-w(x)$, $x\in K$ implies that $\varepsilon^{-1}|x^\varepsilon-y^\varepsilon|^4 \le 2\|u^\kappa\|_\infty+2\|w\|_\infty$, and thus $x^0=y^0$. Moreover, 
\begin{equation}
u^\kappa(x^\varepsilon)-w(y^\varepsilon)
\ge u^\kappa(x^\varepsilon)-w(y^\varepsilon)-\varepsilon^{-1} |x^\varepsilon-y^\varepsilon|^4
\ge u^\kappa(x)-w(x),\quad x\in K
\end{equation}
together with the upper semincontinuity of $u^\kappa$ and $-w$ yield $(u^\kappa-w)(x^0) = \max_K(u-w)$. 
	
\medskip

To obtain (a), it suffices to check that $x^0\in\partial K$. Suppose, on the contrary, that $x^0\in \accentset{\circ}{K}$. Then, $x^\varepsilon,y^\varepsilon\in\accentset{\circ}{K}$ for $\varepsilon>0$ small enough. For such an $\varepsilon$, let $\zeta^\varepsilon(x,y) := -\varepsilon^{-1}|x-y|^4$. If $x^\varepsilon=y^\varepsilon$, then $\nabla_y \zeta^\varepsilon(x^\varepsilon, y^\varepsilon) = 0$ and $\nabla^2_y \zeta^\varepsilon(x^\varepsilon, y^\varepsilon) = 0$. However, $y^\varepsilon$ minimizes  $y\mapsto w(y) - \zeta^\varepsilon(x^\varepsilon, y)$ over $y \in K$. By the supersolution property of $w$ at $y^\varepsilon$, it follows that $1\le F^*(0,0)=0$ (recall Lemma \ref{compute F}), a contradiction. Therefore, $x^\varepsilon \neq y^\varepsilon$.
	
\medskip
	
Since $x^\varepsilon \neq y^\varepsilon$ belong to $\accentset{\circ}{K}$, the Crandall-Ishii Lemma (see \cite{CraIsh}) yields $M^\varepsilon,N^\varepsilon\in\mathbb{S}^n$ with $M^\varepsilon \preceq N^\varepsilon$, $F_*(p^\varepsilon, M^\varepsilon) \le \kappa$ and $F^*(p^\varepsilon, N^\varepsilon) \geq 1$ where $p^\varepsilon := -\nabla_x \zeta^\varepsilon(x^\varepsilon, y^\varepsilon)\neq 0$. By the continuity of $F$ on $({\mathbb R}^n\backslash\{0\})\times\mathbb{S}^n$ (Lemma \ref{compute F}) and its ellipticity, we have:
\begin{equation}\label{Crandall Ishii contr}
\kappa \ge F_*(p^\varepsilon, M^\varepsilon) = F(p^\varepsilon, M^\varepsilon) \ge F(p^\varepsilon, N^\varepsilon) = F^*(p^\varepsilon, N^\varepsilon) \geq 1,
\end{equation}
which contradicts $\kappa\in(0,1)$. This contradiction proves that $x^0 \in \partial K$.
	
\medskip
	
To see (b), we recall that $u^\kappa(x)-w(x)\le\Phi^\varepsilon(x^\varepsilon,y^\varepsilon)$, $x\in K$, $\varepsilon>0$. Moreover, for $\varepsilon>0$ small enough: $y^\varepsilon\in\accentset{\circ}{K'}$ (because $y^\varepsilon\to x^0\in\partial K\subset\accentset{\circ}{K'}$); $x^\varepsilon\neq y^\varepsilon$ by the same contradiction argument as above; and if $u^\kappa$ has the subsolution property at $x^\varepsilon$, then the Crandall-Ishii Lemma yields the contradiction \eqref{Crandall Ishii contr}, so $x^\varepsilon\in\partial K$ and $u^\kappa(x^\varepsilon)\le 0$. In addition, $w\ge0$ on $K'$, as a minimizer $y$ of $w$ with $w(y)<0$ is impossible in view of the supersolution property of $w$ at $y$ (take $\varphi\equiv0$ and recall that $F^*(0,0)=0$). All in all, $\Phi^\varepsilon(x^\varepsilon,y^\varepsilon)=u^\kappa(x^\varepsilon) - w(y^\varepsilon) - \varepsilon^{-1} |x^\varepsilon-y^\varepsilon|^4\le 0$ for $\varepsilon>0$ small enough. \ep

\begin{thm}[Comparison Principle] \label{thm: comparison}
In the setting of Theorem \ref{thm: vis}, suppose that there are $T_\iota\!:\rr^n\to\rr^n$, $\iota\in(1,2]$, each given by a composition of a rotation, a dilation and a translation, and satisfying $K\subset \accentset{\circ}{T_\iota(K)}$, for which $\lim_{\iota\downarrow1} T_\iota=I$. Then, for any upper semicontinuous viscosity subsolution $u$ of $F(\nabla u,\nabla^2u)=1$ on $K$ satisfying the zero boundary condition and any lower semicontinuous viscosity supersolution $w$ of $F(\nabla w,\nabla^2w)=1$ on $K$ satisfying the zero boundary condition, it holds $u\le w^*$.
\end{thm} 

\noindent\textbf{Proof.} Consider $w^\iota\!:T_\iota(K)\to\rr$, $x\mapsto w(T_\iota^{-1}x)$. We aim to show that $c_\iota^2 w^\iota$ is a lower semicontinuous viscosity supersolution of $F(\nabla w,\nabla^2w)=1$ on $T_\iota(K)$ satisfying the zero boundary condition, where $c_\iota>0$ is the dilation factor in $T_\iota$. To this end, we claim that for any orthogonal $n\times n$ matrix $O$ and for all $(p,M)\in\rr^n\times\mathbb{S}^n$: 
\begin{equation} \label{F transf}
F(p, M) = c_\iota^2\, F(O^\top p,\, c_\iota^{-2} O^{\top} M O).
\end{equation}
Indeed, let $\overline{a}\succeq 0$ with $\overline{a}O^\top p=0$,
$\lambda_{(n-k)}(\overline{a})\geq 1$ and $\lambda_{(1)}(\overline{a})\leq L$. Then, $a:=O\overline{a}O^\top\succeq0$ satisfies $ap=0$, $\lambda_{(n-k)}(a)\geq 1$ and $\lambda_{(1)}(a)\leq L$. Thus, the definition of $F$ in \eqref{def: F} yields
\begin{equation} \label{F transf ineq}
F(p, M) \le - \frac{1}{2} \mathrm{tr}(Ma) = - \frac{c_\iota^2}{2}\,\mathrm{tr}(c_\iota^{-2}O^\top M O\overline{a}).
\end{equation}
Taking the infimum over all $\overline{a}$ as described, we arrive at \eqref{F transf} with ``$\le$''. Conversely, we can use the obtained inequality with $c_\iota$, $O$, and $(p,M)$ replaced by $c_\iota^{-1}$, $O^\top$, and $(O^{\top} p,\, c_\iota^{-2} O^{\top} M O)$, respectively, to find that
\begin{equation}
F(O^{\top} p,\, c_\iota^{-2} O^{\top} M O) 
\le c_\iota^{-2}\, F(O O^{\top} p,\, 
c_\iota^2 Oc_\iota^{-2} O^{\top} M O O^\top)=c_\iota^{-2}\, F(p, M).
\end{equation}

\smallskip

For any test function $\varphi\in C^2(\rr^n)$, the function $\varphi^\iota\!:\rr^n\to\rr$, $x\mapsto c_\iota^2\,\varphi(c_\iota^{-1}Ox)$ belongs to $C^2(\rr^n)$ and satisfies $\nabla\varphi^\iota(x)=c_\iota O^\top\nabla\varphi(c_\iota^{-1}Ox)$, $\nabla^2\varphi^\iota(x)=O^\top\nabla^2\varphi(c_\iota^{-1}Ox)O$. These formulas, the definition of $F$ in \eqref{def: F}, and \eqref{F transf} let us conclude that
\begin{equation}
\begin{split}
&\; F(\nabla\varphi^\iota(x),\nabla^2\varphi^\iota(x))
=F(c_\iota O^\top\nabla\varphi(c_\iota^{-1}Ox),\,O^\top\nabla^2\varphi(c_\iota^{-1}Ox)O) \\
& =c_\iota^2\,
F(O^\top\nabla\varphi(c_\iota^{-1}Ox),\,c_\iota^{-2}O^\top\nabla^2\varphi(c_\iota^{-1}Ox)O)
=F(\nabla\varphi(c_\iota^{-1}Ox),\nabla^2\varphi(c_\iota^{-1}Ox)).
\end{split}
\end{equation}
Since the same transformation rule then also holds for $F^*$, the lower semicontinuous $c_\iota^2w^\iota$ is a viscosity supersolution of $F(\nabla w,\nabla^2w)=1$ on $T_\iota(K)$ satisfying the zero boundary condition.

\medskip

Applying Theorem \ref{thm: maximum}(b) we obtain $u\le c_\iota^2w^\iota$ on $K$. Finally, passing to the limit $\iota\downarrow1$ we end up with $u(x)\le\liminf_{\iota\downarrow1} w^\iota(x)=\liminf_{\iota\downarrow1} w(T_\iota^{-1}x)\le w^*(x)$, $x\in K$. \ep

\medskip

We are now ready for the proof of Proposition \ref{prop: uniq}.

\medskip

\noindent\textbf{Proof of Proposition \ref{prop: uniq}.} Let $v$, $\widetilde{v}$ be upper semicontinuous viscosity solutions of $F(\nabla v,\nabla^2v)=1$ on $K$ satisfying the zero boundary condition. By Theorem \ref{thm: comparison} with $u:=v$ and $w:=\widetilde{v}_*$, it holds $v\le(\widetilde{v}_*)^*\le\widetilde{v}^*=\widetilde{v}$. For the same reason, $\widetilde{v}\le v$. \ep

\section{Continuity of the Value Function} \label{sec5}

In this final section, we discuss continuity properties of the value function $v$ from \eqref{soc problem} assuming that the compact $K\subset\rr^n$ is \textit{convex}. We start with two simple observations. 

\begin{prop} \label{lem:continuity_intK}
Let the compact $K\subset\mathbb{R}^n$ be convex. Then, the value function $v$ of \eqref{soc problem} is continuous on $\accentset{\circ}{K}$.
\end{prop}

\noindent\textbf{Proof.} Consider an $x\in \accentset{\circ}{K}$. For $\iota\in(0,1)$, we define $T_\iota\!: K\to\rr^n$, $y\mapsto x+\iota(y-x)$. Since $x\in \accentset{\circ}{K}$ and $K$ is convex, $T_\iota(K)\subset\accentset{\circ}{K}$. By arguing as in the proof of Theorem \ref{thm: comparison}, we find that $v^\iota\!:T_\iota(K)\to\rr$, $y\mapsto \iota^2v(T_\iota^{-1}y)$ is an upper semicontinuous viscosity solution of $F(\nabla v,\nabla^2 v)=1$ on $T_\iota(K)$ satisfying the zero boundary condition. Thus, $v^\iota\le v_*$ on $T_\iota(K)$ by Theorem \ref{thm: maximum}(b). In particular, for $x=T_\iota x\in T_\iota(K)$, we have $v_*(x)\ge v^\iota(x)=\iota^2 v(x)$. Taking the limit $\iota\uparrow1$ we arrive at $v_*(x)\ge v(x)=v^*(x)$. Hence, $v$ is continuous on $\accentset{\circ}{K}$. \ep

\begin{prop}
Let the compact $K\subset\mathbb{R}^n$ be convex. If the value function $v$ of \eqref{soc problem} satisfies $v\equiv 0$ on $\partial K$, then $v$ is continuous on $K$.  
\end{prop}

\noindent\textbf{Proof.} Since $v$ is continuous on $\accentset{\circ}{K}$ by Proposition \ref{lem:continuity_intK}, it suffices to consider points $x\in\partial K$. By the definition of $v$ in \eqref{soc problem}, $v\ge0$, and so $v_*\ge0$. Together with the upper semicontinuity of $v$ (Proposition \ref{prop: dpp}), we get 
\begin{equation}\label{0=>cont}
0=v(x)=v^*(x)\ge v_*(x)\ge0, 
\end{equation}
i.e., $v^*(x)=v_*(x)=0$. \ep

\medskip

More generally, the boundary behavior of $v$ is characterized in the next lemma.

\begin{lemma} \label{lem: v=0}
Let the compact $K\subset\mathbb{R}^n$ be convex. For $x\in K$, it holds $v(x) = 0$ if and only if $\text{dim}(F_x)\leq n-k$, where $F_x$ is the unique face of $K$ whose relative interior contains $x$. 
\end{lemma}

\noindent\textbf{Proof.} For simplicity of notation, we assume that $x=0$, which can be achieved by a translation. Define $Q$ as the orthogonal projection onto the orthogonal complement of $F_0$ and write $X$ for the coordinate process under an optimal $\mathrm{P} \in \mathcal{P}_0$. Next, we distinguish three cases. If $\text{dim}(F_0) < n-k$, then on the one hand, $Q\,\mathrm{d}X(t)=0$, $t\le\tau_{F_0}$, and consequently $0=\text{tr}(\mathrm{d}\langle QX\rangle(t))=\text{tr}(Q\,\mathrm{d}\langle X\rangle(t))$, $t\le\tau_{F_0}$. On the other hand, by the first line in \eqref{eq: constraint}, $\text{tr}\big(Q\,\frac{\mathrm{d}\langle X\rangle(t)}{\mathrm{d}t}\big)\ge n-\text{dim}(F_0)-k$, a.e. $t\ge0$. If $\text{dim}(F_0)<n-k$, we find that $\tau_{F_0}=0$. Repeating \cite[proof of Lemma 5.2]{larsson_minimum_2022} word by word we conclude that $\tau_K=\tau_{F_0}=0$.
	
\medskip
	
If $\text{dim}(F_0) = n-k$, we consider an open ball $B_r(0)$ in the linear subspace spanned by $F_0$ such that $F_0\subset B_r(0)$. Then, for any $q\in B_r(0)\backslash\{0\}$, the process $\frac{q^\top X(t)}{|q|}$, $t\le\tau_{F_0}$ is a stopped sped-up standard Brownian motion due to the first line in \eqref{eq: constraint}. Since the exit time of the latter from $[-r,r]$ has an essential infimum of $0$, we have $0=\mathrm{P}\text{-ess} \inf\tau_{F_0}=\mathrm{P}\text{-ess} \inf\tau_K$.
	
\medskip
	
If $\text{dim}(F_0)>n-k$, then there exists a non-trivial closed ball $\overline{B_r(0)}$ in the linear subspace spanned by $F_0$ such that $\overline{B_r(0)}\subset F_0$. Recalling the measure $\mathrm{P}^{\ast}\in\mathcal{P}_0$ from Example \ref{ex: ball}, we obtain $v(0)\ge \mathrm{P}^*\text{-ess} \inf\tau_K=\mathrm{P}^*\text{-ess} \inf\tau_{F_0}\ge\mathrm{P}^*\text{-ess} \inf\tau_{\overline{B_r(0)}}=r^2/(n-k)>0$. \ep

\medskip

Lastly, Lemma \ref{lem: v=0} leads to the following two propositions. 

\begin{prop}
Let the compact $K\subset\mathbb{R}^n$ be convex. If $k=1$ or $k=2$, then the value function $v$ of \eqref{soc problem} is continuous on $K$.
\end{prop}

\noindent\textbf{Proof.} It suffices to combine Lemma \ref{lem: v=0} with \cite[Lemmas 5.7 and 5.6]{larsson_minimum_2022}, whose proofs can be repeated word by word. \ep

\begin{prop}
Let $K\subset\rr^n$ be a polytope. Then, the value function $v$ of \eqref{soc problem} is continuous on $K$.
\end{prop}

\noindent\textbf{Proof.} We can repeat \cite[proof of Corollary 5.9(iii)]{larsson_minimum_2022} word by word, only using our Lemma \ref{lem: v=0} instead of their Lemma 5.3. \ep

\bigskip\bigskip

\bibliographystyle{amsalpha}
\bibliography{reference}

\bigskip\bigskip

\end{document}